\documentclass[10pt,draft]{dis03}
\usepackage{epsf,amsmath}

\newcommand{\beq} {\begin{equation}}
\newcommand{\eeq} {\end{equation}}
\newcommand{\beqa} {\begin{eqnarray}}
\newcommand{\eeqa} {\end{eqnarray}}

\newcommand{\ie}{{\it i.e.}}

\newcommand{\etal}{{\it et al.}}

\newcommand{\im}{{\rm Im}}

\newcommand{\qu}{{\rm q}}

\newcommand{\qbm}{{\rm\bar q}}

\newcommand{\rvec}{\vec r}
\newcommand{\Rvec}{\vec R}

\newcommand{\pl}{{||}}

\newcommand{\morder}[1]{{\cal O}\left(#1 \right)}
\newcommand{\eq}[1]{(\ref{#1})}

\newcommand{\ave}[1]{\langle{#1}\rangle}

\begin{document}

\title{
\hbox to\hsize{\normalsize\hfil\rm LAPTH-Conf-991/03}
\hbox to\hsize{\normalsize\hfil hep-ph/0307261}
\hbox to\hsize{\normalsize\hfil July 21, 2003}
\vskip 30pt
\centerline{UNIVERSALITY-BREAKING EFFECTS}
\centerline{IN DIS AND DRELL-YAN\footnote{To appear in the proceedings
of the XIth International Workshop on Deep Inelastic Scattering,  
St. Petersburg, 23-27 April 2003.}}} 
\author{S.~Peign\'e \\
LAPTH, B.P. 110, F-74941 Annecy-le-Vieux Cedex, France\\
E-mail: peigne@lapp.in2p3.fr}
\maketitle

\begin{abstract}
\noindent Several properties of high energy hadronic
collisions are illustrated by comparing DIS and the Drell-Yan process
within a scalar QED model. 
Diffraction and transverse momentum broadening within the target system
are found to be non-universal.
Here these effects are simply due to the Coulomb phase shift
in the Drell-Yan production amplitude.
\end{abstract}

\section{Introduction and summary} 
In high-energy hadron-hadron collisions, QCD factorization theorems
valid for leading-twist inclusive cross sections are known to fail 
in hard diffraction. For instance, in diffractive jet production, 
$A+B \rightarrow A + \rm{jets} + X$, there exist non-factorizable
leading-twist contributions to the cross section \cite{Berera:1994xh} .
Those contributions arise when the second hard gluon exchange,
necessary to leave hadron $A$ intact, occurs either with the active 
parton of the hard jet production subprocess, or with a spectator 
parton of the projectile hadron $B$. This type of process, called
`coherent hard diffraction' \cite{Collins:cv},
cannot be expressed in terms of {\it universal}
diffractive parton distributions and thus breaks factorization.

In general, observables which are 
{\it differential} in target fragmentation can violate universality.
One way to break universality is to constrain the target to be 
diffractively scattered. Another is to measure the transverse momentum
between some of the target fragments. 
In this talk I present a model comparing DIS
and Drell-Yan (DY), where both universality-breaking effects
are due to the infrared sensitive 
Coulomb phase factor in the DY production amplitude. 
This factor leads to a suppression of diffractive events and to less
transverse momentum broadening in DY compared to DIS.

\section{Model for DIS and DY parton distributions}
The parton distributions are modelled within scalar QED,
the simplest gauge theory where factorization and
universality can be tested. For the DIS forward Compton 
amplitude we use the model of Ref.~\cite{BHMPS} which is pictured 
in Fig.~1a. The target `quark' is a scalar of mass $M$
and the `quark' and `antiquark' of momenta $p_1$ and $p_2$
scalars of mass $m$.
The model for DY production \cite{peigne} is directly
obtained (Fig.~1b) by crossing the lines of the virtual photon 
$\gamma^*(q)$ and of the struck scalar `quark' $\qu(p_1)$ , 
and by the replacement
$q^2=-Q^2 <0 \rightarrow q^2= Q^2 >0$. 
We work in a target rest frame where\footnote{We use the 
light-cone variables $k^{\pm}=k^0\pm k^z$ to define a momentum
$k=(k^+, k^-, \vec{k}_{\perp})$.} $q=(\mp Mx_B, q^-, \vec{0}_{\perp})$ 
for DIS and DY respectively, with $q^- \equiv 2\nu = Q^2/Mx_B$.
\begin{figure}[h]
\vspace*{5.5cm}
\begin{center}
\includegraphics{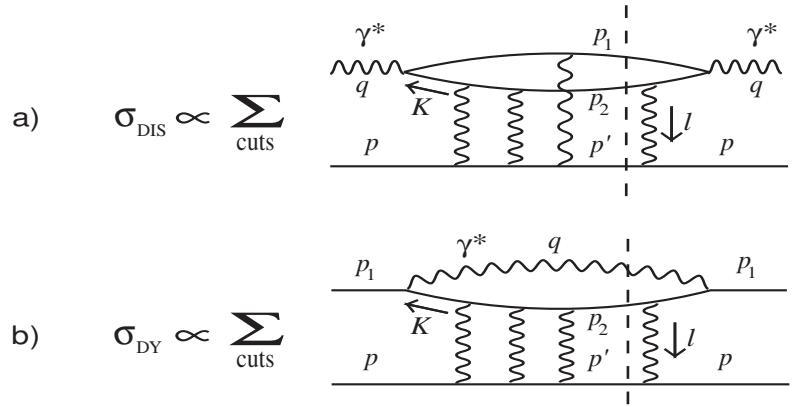}
\caption[*]{SQED model for the quark distribution in 
DIS (a) and DY (b).}
\end{center}
\end{figure}

\vskip -5mm
We briefly recall the main features of the model.
Since one concentrates on Coulomb rescatterings, 
extracting in the Bjorken limit
the leading-twist cross section requires 
focussing on the aligned jet kinematical region. Thus most of the 
incoming energy $\nu$ is transferred to the 
struck scalar quark in DIS (Fig.~1a) or to the outgoing virtual photon in 
DY (Fig.~1b), \ie\ $p_1^- \simeq q^-$. Note that in the chosen Lorentz
frame the momentum $K$ (see Fig.~1) satisfies $K^+>0$, giving for the   
hard subprocess $\gamma^* \qu \rightarrow \qu$ in DIS and
$\qbm \qu \rightarrow \gamma^*$ in DY.
The hard vertex is given by the bare
virtual photon coupling $e$, \ie\ the hard scale
$\nu$ does not flow in internal propagators. Thus the contribution
to $\sigma_{DIS}$ or $\sigma_{DY}$ we calculate is directly interpreted
as a contribution to the scalar quark distribution in the target
$f_{\qu/T}(x)$ probed in DIS or DY, and evaluated at $x=x_B$\footnote{Indeed 
$x=K^+/p^+\simeq \mp q^+/p^+=x_B$ in DIS and DY respectively.}. 
Diagrams contributing to the $Q^2$ evolution of $f_{\qu/T}$ are not
included. Our model thus describes $f_{\qu/T}(x,Q_0)$ at an initial
soft scale $Q_0$. This is sufficient to study questions related to 
universality. 
  
The features I will discuss in the next sections are most easily
obtained after resumming Coulomb exchanges. This resummation can be
done explicitly in the limit $x_B \ll 1$. 
In transverse coordinate space the
production amplitudes 
${\cal M}_{DIS}(\gamma^*T \rightarrow \qu(p_1)\qbm(p_2)T(p'))$ 
and 
${\cal M}_{DY}(\qbm(p_1)T \rightarrow \gamma^* \qbm(p_2)T(p'))$ 
are simply related by a phase factor
(see Eq.~(39) of Ref.~\cite{BHMPS} and Eq.~(B6) of Ref.~\cite{peigne}):
\beqa
{\cal M}_{DY}(\vec{r}_{\perp},\vec{R}_{\perp}) &=& 
- \,  e^{i \phi_{\lambda}(R_\perp)} 
{\cal M}_{DIS}(\vec{r}_{\perp},\vec{R}_{\perp}) 
\label{phase} \\
\phi_{\lambda}(R_\perp) &=& g^2 \int \frac{d^2\vec{l}_\perp}{(2\pi)^2}
\frac{e^{i\vec{R}_\perp\cdot\vec{l}_{\perp}}}{l_\perp^2 + \lambda^2 } 
= \frac{g^2}{2\pi} \, K_0(\lambda R_\perp)
\label{phi} 
\eeqa
For further convenience we give \cite{BHMPS,peigne}:
\beqa
{\cal M}_{DIS}(\vec{r}_{\perp},\vec{R}_{\perp}) &=& Mp_2^- \,
\psi_{\gamma^*}(r_{\perp})\ T_{\qu \qbm}(\vec{r}_\perp, \vec{R}_\perp) 
\label{wfdip} \\
\psi_{\gamma^*}(r_{\perp}) &=& \frac{eQ}{\pi} \, K_0(m_{\pl} r_\perp)
\label{wf} \\
m_{\pl}^2 &=& p_2^- Mx_B + m^2 \\
T_{\qu \qbm}(\vec{r}_\perp, \vec{R}_\perp) &=&   
2 \sin(W_{\lambda}/2)\  e^{-i W_{\lambda}/2} 
\label{T} \\
W_{\lambda}(\vec{r}_\perp, \vec{R}_\perp) &=& \phi_{\lambda}(R_\perp) -
\phi_{\lambda}(|\vec{R}_\perp + \vec{r}_{\perp}|) 
\label{W} 
\eeqa
In the above equations $\vec{r}_{\perp}$ and $\vec{R}_{\perp}$ 
are the variables conjugate to 
$\vec{p}_{2\perp}$ and $\vec{l}_{\perp}$, where $l=p-p'$ is the total 
Coulomb momentum exchange, $g$ denotes the
coupling of the exchanged Coulomb photons to the scalar quarks,
and we introduced a finite photon mass $\lambda$ as an infrared 
regulator. 

From Eq.~\eq{phase} the DIS and DY cross sections 
integrated over $\vec{r}_{\perp}$ and $\vec{R}_{\perp}$
are identical. Our model is thus consistent with the universality
of the inclusive scalar quark distribution, namely
$f_{\qu/T}^{DIS}(x)=f_{\qu/T}^{DY}(x)$. 
In particular leading-twist shadowing
effects are the same in DIS and DY. One can easily show that the 
$K_{\perp}$-dependent distribution is also universal, 
$f_{\qu/T}^{DIS}(x,K_{\perp})=f_{\qu/T}^{DY}(x,K_{\perp})$.

\section{Violation of universality in diffractive events}
The transverse coordinate space DIS and DY amplitudes are identical 
{\it up to the phase shift} $\phi_{\lambda}(R_\perp)$,
which arises because the DY
process involves the scattering of a charge instead of a dipole in
DIS. In Eq.~\eq{wfdip} $\psi_{\gamma^*}(r_{\perp})$ is the 
$\gamma^* \to \qu \qbm$ wavefunction and 
$T_{\qu\qbm}(\vec{r}_\perp, \vec{R}_\perp)$ the $\qu \qbm$
dipole scattering amplitude. We recover the expression of
$\sigma_{DIS}$ in terms of the $\qu \qbm$ dipole cross section
$\sigma_{\qu \qbm}$ \cite{zn}:
\beqa
\sigma_{DIS} &=& \sigma_{DY} \propto \int d^2\vec{r}_{\perp} \, 
|\psi_{\gamma^*}(r_{\perp})|^2 \, \sigma_{\qu \qbm}(r_{\perp})
\label{dipolesigmaDIS} \\
\sigma_{\qu \qbm}(r_{\perp}) &=& \int d^2\vec{R}_{\perp} \, 
|T_{\qu\qbm}(\vec{r}_{\perp}, \vec{R}_{\perp})|^2
\label{dipole}
\eeqa

In our model, the diffractive cross
section is identified with $C$-even exchanges, 
which correspond (for $x_B \ll 1$), to the 
imaginary part of the production amplitude. In DIS the  
$|t| = l_{\perp}^2 = 0$ diffractive cross section reads:
\beqa
\left. \frac{d\sigma^{diff}_{DIS}}{d l_{\perp}^2} \right)_{l_{\perp}^2 =0} 
&\propto& \int d^2\rvec_{\perp} |\psi_{\gamma^*}(r_{\perp})|^2 
\left| \int d^2\Rvec_{\perp} \, \im\, T_{\qu \qbm} \right|^2 
\label{diffDISt0} \\
&\propto&  \int d^2\rvec_{\perp} |\psi_{\gamma^*}(r_{\perp})|^2 \,
\sigma_{\qu \qbm}(r_{\perp})^2
\label{dipolediffDISt0}
\eeqa
where the second line is obtained from the unitarity relation
following from \eq{T}:
\beq
2 \im \,T_{\qu\qbm}  = - 4 \sin^2(W_{\lambda}/2) 
=  - |T_{\qu\qbm}|^2 
\label{unitarity}
\eeq
Comparing \eq{dipolesigmaDIS} to \eq{dipolediffDISt0} one observes
the close relationship between the total and $t = 0$ diffractive DIS
cross sections \cite{zn}. 

From \eq{phase} this relation does not hold in the DY case, as can be
directly seen by replacing in \eq{diffDISt0} $\im\, T_{\qu \qbm}$ 
by 
\beq
\im\, T_{DY} = 
\im\,\left[ - e^{i\phi_{\lambda}(R_\perp)}\ T_{\qu \qbm} \right] \neq
\im\, T_{\qu \qbm}
\label{TDY}
\eeq
We easily obtain from \eq{T},  \eq{W} and \eq{TDY}:
\beq
\int d^2\Rvec_{\perp} \, \im\, T_{DY} = \int d^2\Rvec_{\perp} \,
\left\{  \cos[\phi_{\lambda}(R_\perp)] - 
\cos[\phi_{\lambda}(|\Rvec_\perp+\rvec_\perp|)] \right\} = 0 
\eeq

One thus finds that at leading-twist the $t = 0$ 
diffractive DY cross section vanishes. As a consequence 
the {\it diffractive}
scalar quark distribution is non-universal. This is due to 
the presence of the Coulomb phase in \eq{TDY}, which  
spoils the DIS unitarity relation \eq{unitarity}.

\section{Violation of universality in momentum-broadening}
Now we fix the total transverse momentum exchange 
$l_{\perp}$. Since the soft quark of momentum $p_2$ 
belongs to the outgoing target system (see Fig.~1), 
$l_{\perp}$ is a variable internal
to the target structure, different from the probed transverse 
momentum $K_{\perp}$. We give below a heuristic argument why fixing 
$l_{\perp}$ breaks universality and refer to Ref.~\cite{peigne}
for a proof.

Remembering that $\vec{l}_{\perp}$ is conjugate to $\vec{R}_{\perp}$
we have, using \eq{phase}:
\beq
{\tilde {\cal M}}_{DY}(\vec{r}_{\perp},\vec{l}_{\perp}) 
= - \int d^2\vec{R}_\perp 
e^{-i\vec{R}_{\perp}\cdot \vec{l}_{\perp} + i \phi_{\lambda}(R_\perp)}
\, {\cal M}_{DIS}(\vec{r}_{\perp},\vec{R}_{\perp})
\label{mixed}
\eeq
At fixed $\vec{l}_{\perp}$, ${\tilde {\cal M}}_{DY}$ is 
suppressed because the infrared sensitivity of 
$\phi_{\lambda}(R_\perp)$ makes the 
phase factor rapidly oscillating. Indeed, when $\lambda \to 0$, 
$R_\perp \sim 1/\lambda \to \infty$. Strictly speaking, since 
$\phi_{\lambda}(R_\perp) \propto g^2$, this 
heuristic argument holds only beyond leading order.
In Ref.~\cite{peigne} the $\morder{g^4}$ shadowing 
correction to the Born DY cross section is calculated, and 
the typical values of $l_{\perp}$ contributing to 
$\sigma_{DY}-\sigma_{DY}^{Born}$ are indeed shown to scale with
$\lambda$:
\beq
\ave{l_\perp}_{DY} \sim \lambda \ll \ave{l_\perp}_{DIS} \sim m
\eeq
Thus leading-twist transverse momentum broadening within the target
is non-universal, and is suppressed in DY compared to
DIS in the present model. 

\section*{Acknowledgements} 
This talk is based on discussions with S.~J.~Brodsky, 
P.~Hoyer and D.~S.~Hwang.

\end{document}